\documentstyle[12pt,epsf]{article}
\catcode`\@=11
\def \thesection {\arabic{section}.}

\def \be  {\begin{equation}}
\def \ee  {\end{equation}}
\def \ba  {\begin{eqnarray}}
\def \ea  {\end{eqnarray}}
\def \baa {\begin{eqnarray*}}
\def \eaa {\end{eqnarray*}}
\def \bb  {\begin{thebibliography}}
\def \eb  {\end{thebibliography}}
\def \lab #1 {\label{#1}}
\newcommand \ci [1] {\cite{#1}}
\newcommand \bi [1] {\bibitem{#1}}
\newcommand\re[1]{(\ref{#1})}

\def \matrix #1 {\left(\begin{array}{cc} #1 \end{array}\right)}

\def \e  {\mathop{\rm e}\nolimits}
\def \PT {{_{\rm PT}}}
\def \QCD {{_{\rm QCD}}}
\def \MS {\overline{_{\rm MS}}}

\newcommand\lr[1]{{\left({#1}\right)}}

\newcommand \vev [1] {\langle{#1}\rangle}

\newcommand \ket [1] {|{#1}\rangle}

\newcommand{\as}{\ifmmode\alpha_{\rm s}\else{$\alpha_{\rm s}$}\fi}

\def \CO {{\cal O}}

\font\cmss=cmss12 
\def\inbar{\,\vrule height1.5ex width.4pt depth0pt}
\def\IC{\relax\hbox{$\inbar\kern-.3em{\rm C}$}}
\def\IZ{\relax{\hbox{\cmss Z\kern-.4em Z}}}
\def\IR{{\hbox{{\rm I}\kern-.2em\hbox{\rm R}}}}
\def\IP{{\hbox{{\rm I}\kern-.2em\hbox{\rm P}}}}
\def\II{\hbox{{1}\kern-.25em\hbox{l}}}

\def\IK{{\hbox{{\rm I}\kern-.2em\hbox{\rm K}}}}

\def\numberbysection{\@addtoreset{equation}{section}
                     \def\theequation{\thesection\arabic{equation}}}

\begin{document}

\def\thefootnote{\fnsymbol{footnote}}
\thispagestyle{empty}

\hfill\parbox{50mm}{{\sc LPTHE--Orsay--98--44} \par
                         hep-ph/9806537       \par
                         June, 1998}
\vspace*{35mm}
\begin{center}
{\LARGE Shape functions and power corrections to the event shapes}
\footnote{Talks presented at the 33rd Rencontres de Moriond 
``QCD and High Energy Hadronic Interactions'', March 21-28, 1998, Les Arcs, 
France and the 3rd Workshop on ``Continuous Advances in QCD'', April 16-19, 1998,
University of Minnesota, Minneapolis}
\par\vspace*{15mm}\par
{\large G.~P.~Korchemsky}
\footnote{Also at the Laboratory of Theoretical Physics, 
Joint Institute for Nuclear Research, 141980 Dubna, Russia}

\par\bigskip\par\medskip

{\em Laboratoire de Physique Th\'eorique et Hautes Energies%
\footnote{Laboratoire associ\'e au Centre National de la Recherche
Scientifique (URA D063)} \\
Universit\'e de Paris XI, Centre d'Orsay, b\^at. 210\\
91405 Orsay C\'edex, France}
\end{center}

\vspace*{20mm}

\begin{abstract}
We show that the leading power corrections to the event shape distributions 
can be resummed into nonperturbative shape functions that do not depend on 
the center-of-mass energy and measure the energy flow in the final state. 
In the case of the thrust variable, the distribution is given by the 
convolution of the perturbative spectrum with the shape function. Choosing 
the simplest ansatz for the shape function we find that our predictions for 
the thrust distribution provide a good description of the data in a wide 
range of energies.
\end{abstract}

\newpage

\def\thefootnote{\arabic{footnote}}
\setcounter{footnote} 0

\section{Introduction}

Study of hadronization corrections to the event shapes in $\e^+\e^-$
annihilation became a unique laboratory for testing QCD dynamics beyond
perturbative level \ci{evi}. Being infrared and collinear safe quantities, 
the event shapes (their mean values as well as differential distributions) can 
be calculated in perturbative QCD at large center-of-mass energies $s\equiv
Q^2$ as series in $\as(Q)$. Nonperturbative corrections to the event
shapes are attributed to hadronization effects and they are expecting to
modify perturbative predictions by terms suppressed by powers of large
scale $1/Q^p$ with the exponent $p$ varying for different observables.

Successful description of the hadronization effects by the
phenomenological Monte-Carlo based models indicates that in distinction
with the total cross-section of $\e^+\e^-$ annihilation the power
corrections to the event shapes become anomalously large and for the shape
variables like the thrust and heavy mass jet they are expecting to
appear at the level $p=1$.

The enhancement of hadronization corrections occurs due to the fact that
the event shapes are not completely inclusive quantities with respect to
the final states but rather weighted cross-sections in which large power
corrections can be attributed to an incomplete cancellation of the
contribution of soft gluons. As a consequence,
the operator product expansion (OPE) is not applicable to the analysis of
the event shapes and the standard identification of the exponents $p$
characterizing the strength of power corrections as dimensions of
local composite operators entering the OPE does not hold.

To determine the leading exponent $p$ and also to understand the
way in which nonperturbative effects modify perturbative predictions one
may explore instead by now standard infrared renormalon analysis. This
procedure has been successfully applied to the mean value of different event
shapes variables and the description of the leading $1/Q-$power corrections has
been given within different approaches \ci{W}-\ci{DMW}. In contrast, the 
hadronization corrections to the differential event shape distributions are 
less understood. One of the reasons for this is that the leading power 
corrections to the mean values and to the differential distributions have 
different form \ci{evi}: the former are characterized by a single 
nonperturbative scale of dimension $p$ while the latter involve the 
nonperturbative function of the shape variable that one usually estimates 
running the Monte-Carlo event generators.

Studying the power corrections to the event 
shape distributions we will follow the approach proposed in \ci{KS94}. We will 
mostly concentrate on the differential distribution with respect to the thrust
variable $d\sigma/d T$ and, particularly, in the end-point part of the
spectrum $T\sim 1$.

There are few reasons for considering the region $T\sim 1$. In contrast
with the mean value $\vev{1-T}$ that gets $1/Q-$power correction from
the final states with an arbitrary number of jets, for the thrust
distribution in the end-point region, $T\to 1$, one has in the final state 
only two narrow energetic jets moving close to the light-cone directions
$p_+$ and $p_-$ ($p_\pm=\frac{Q}2(1,{\bf 0},\pm 1)$ and $2(p_+p_-)=Q^2$) in 
two opposite hemispheres.  Denoting their invariant masses as $M_L^2$ and 
$M_R^2$ one gets
\be
t\equiv 1-T
\simeq\frac{M_R^2}{Q^2}+\frac{M_L^2}{Q^2}\,,
\lab{t}
\ee
where for later convenience we introduced a new variable $t$.

Taking into account the QCD effects of collinear splitting of
quark and gluons inside two narrow jets and their interaction
with soft gluon radiation one finds that
the thrust distribution for $t\to 0$ depends on two infrared scales,
$Q^2t^2$ and $Q^2 t$, which give rise to large both perturbative (Sudakov)
logs and power corrections. The smallest scale, $(Qt)^2$, sets up the total
energy carried by soft gluons in the final state and the scale $Q^2t$
characterizes the transverse size of the jets $k_\perp^2=\CO(Q^2t)$. The
power corrections to the thrust distribution are suppressed by powers of
both scales. In order to separate the leading asymptotics one keeps the 
smaller scale $Qt$ fixed and expands the thrust distribution in powers
of the larger scale $Q^2t$
\be
\frac1{\sigma_{\rm tot}}\frac{d\sigma}{dt}=
\sigma_0\lr{\as(Q),\ln t,\frac1{Qt}} + \CO\lr{\frac1{Q^2t}}\,.
\lab{s0}
\ee
In what follows we will consider only the leading term of this
expansion, $\sigma_0$. It should resums large perturbative terms
$\as^n(Q)\ln^{m}t/t$ ($m \leq 2n-1$) and take into account all
power corrections of the form $1/(Qt)^n$.

The structure of power corrections $\sim 1/(Qt)^n$ strongly depends on the
value of the thrust variable. Away from the end-point region,
$t\gg\Lambda_\QCD/Q$ one may retain in $\sigma_0$ only the lowest term $n=1$ and
neglect the terms with $n\ge 2$ as suppressed by powers of $1/Q$.
It is this approximation that one applies calculating the mean value of
the thrust
$
\vev{t} \equiv  \sigma_{\rm tot}^{-1}\int_0^{t_{\rm max}} dt\, t
\frac{d\sigma}{dt}\,,
$
where $t_{\rm max}$ is the upper limit on the thrust variable that one 
imposes to separate the contribution of the 2-jet final state, $t_{\rm
max}=\frac13$. At the same time, in the end-point region 
$t=\CO(\Lambda_\QCD/Q)$ all terms $\sim(Qt)^{-n}$ become equally important 
and need to be resummed inside $\sigma_0$ for all $n$.

As we will show, the leading power corrections to
the thrust distribution in the end-point region away from the small invariant
jet mass limit $Q^2t\gg \Lambda_{\QCD}^2$ can be resummed into a
nonperturbative $Q-$independent function that defines the shape of the
distribution in the region $t=\CO(\Lambda_{\QCD}/Q)$ and therefore is called
the shape function. Then, the QCD prediction for the leading term in
\re{s0} is given by the convolution of perturbative Sudakov spectrum with
nonperturbative shape function. One should mention that one finds
similar expressions considering, for example, the large$-x$ asymptotics of
the structure function of deep inelastic scattering \ci{DIS,WL} and the 
end-point spectrum of the inclusive heavy meson decays $B\to \gamma X_s$ 
\ci{BSUV,N,GK}. The reason for this similarity is that in all these cases one 
encounters the same physical situation when energetic narrow jet(s) is 
propagating in the final state through the cloud of soft gluons. However the 
important difference with the thrust distribution is that the nonperturbative 
functions in the latter two cases resum power corrections on the different 
scale $Q^2t$ and they coincide with well-known inclusive (light-cone) 
distributions.

\section{Analysis of soft gluon effects}

At the Born level the final state consists of a quark-antiquark pair and
the thrust distribution has the form
\be
\frac1{\sigma_{\rm tot}}\frac{d\sigma}{dt}\bigg|_{\rm Born}
=\delta(t)\,.
\lab{Born}
\ee
Soft gluon radiation smeares this peak towards larger $t$. Let us first 
analyze separately perturbative and nonperturbative contributions.

Considering the perturbative emissions of soft gluons out of two
outgoing quarks one finds that for $t\to 0$ the phase space for
real soft gluons is squeezed and due to an incomplete cancellation
between virtual and real gluon contributions the perturbative
corrections to the thrust distribution involve large Sudakov logs
that can be resummed to all orders with the double logarithmic (DL) accuracy
as
\be
\frac1{\sigma_{\rm tot}}\frac{d\sigma}{dt}\bigg|_{\rm PT}=
\frac{d R_\PT(t)}{dt}\,,
\qquad R_\PT(t)\stackrel{\tiny \hbox{\rm DL}}{=}
\exp\lr{-\frac{4\as(Q)}{3\pi}\ln^2t}
\lab{PT}
\ee
with $R_\PT(t)$ called the radiator function. One can systematically improve
perturbative approximation by including  additional nonleading logarithmic
terms in $R_\PT(t)$ and matching the result into exact higher order
calculations using the $\ln R-$scheme \ci{CTTW}. The perturbative Sudakov
spectrum extends over the interval $0\le t \le t_{\rm max}$ and vanishes at the
end points. The peak of the distribution is located close to $t=0$ and it is
shifted towards larger $t$ as one improves perturbative approximation. Its
position, $t_{\rm p}=\CO(\Lambda_\QCD/Q)$, is sensitive to the emission of
soft gluons with energy $\sim \Lambda_{\QCD}$ indicating that the physical
spectrum around the peak is of nonperturbative origin.

Let us now estimate the effects of nonperturbative soft gluon emissions
on the thrust distribution \re{Born}. We take into account that in the
leading order in $1/(Q^2t)$ the transverse size of two quark jets
$k_\perp^2=\CO(Q^2t)$ can be neglected, that is soft gluons with the energy
$\sim Qt$ can not resolve the internal structure of jets. This means that
considering soft gluon emissions we may apply the eikonal approximation
and effectively replace quark jets by two relativistic classical particles
that carry the color charges of quarks and move apart along the light-cone
directions $p_+$ and $p_-$. The interaction of the quark jets with soft
gluons is factorized into the unitary eikonal phase $W(0)$ given by the
product of two Wilson lines calculated along classical trajectories of two
particles
\be
W(0)=W_+(0)
[W_-(0)]^\dagger\,,\qquad W_\pm(x)=P\exp\lr{i\int_0^\infty ds\, p_\pm\cdot
A(x+p_\pm s)}\,,
\lab{W}
\ee
with gauge fields $A_\mu(x)$ describing soft gluons. Denoting the total
momentum of soft gluons emitted into the right and left hemispheres
as $k_R=\sum_{i\in R}k_i$ and $k_L=\sum_{i\in L}k_i$, correspondingly, one
finds the thrust \re{t} as $t=2(k_Rp_+)/Q^2+2(k_Lp_-)/Q^2$ and obtains the
following expression for the differential distribution
\be
\frac1{\sigma_{\rm
tot}}\frac{d\sigma}{dt}= \sum_N \left|\langle 0|W(0)|N\rangle\right|^2
\delta\lr{t-\frac{k_R^-}{Q}-\frac{k_L^+}{Q}}
\lab{fact}
\ee
with $k^\pm=k_0\pm k_3$. Here, the matrix element of the Wilson line
operator describes the interaction of quarks with soft gluons and
the summation goes over the final states $N$ of soft gluons
with the total momentum $k=k_R+k_L$. Expression \re{fact} follows from
the universality of soft gluon radiation and it takes into account both
perturbative and nonperturbative corrections \ci{WL}.

Let us neglect for the moment the perturbative contribution to the
matrix element of the Wilson line in \re{fact}. Then, introducing the
{\it shape\/} function
\be
f(\varepsilon)=\sum_N \left|\langle 0|W(0)|N\rangle\right|^2
\delta\lr{\varepsilon-k_R^--k_L^+}
\lab{f}
\ee
one can estimate the nonperturbative contribution to the thrust
distribution as
\be
\frac1{\sigma_{\rm tot}}\frac{d\sigma}{dt}\bigg|_{\rm nonPT}=Qf(Qt)\,.
\lab{nonPT}
\ee
The nonperturbative function $f(\varepsilon)$ is localized at small
energies $\varepsilon$ and according to \re{nonPT} it determines the
shape of the spectrum at small $t=\CO(\Lambda_\QCD/Q)$.

The important property of the function \re{f} is that it does not depend on
the center-of-mass energy $Q$. Although the vectors $p_\pm$ entering
into the definition \re{W} of $W(0)$ do depend on $Q$, the $Q-$dependence of
the Wilson lines \re{W} disappears due to the reparameterization invariance
$s\to \lambda s$. Therefore, one may extract the shape function at some
reference $Q_0$ and then apply it to describe the hadronization effects for
different center-of-mass energy.

Being a new nonperturbative distribution, the shape function \re{f} admits the
operator definition that is different however from the one for the inclusive
distributions. As a manifestation of noninclusiveness of the thrust variable,
$f(\varepsilon)$ depends separately on soft gluon momenta flowing into different
hemispheres. In particular, it takes into account the effects of soft gluon
splittings when the decay products fly into two different hemispheres \ci{NS}.
Since this may happen at time scales larger then $1/Q$, we do not expect the
function $f(\varepsilon)$ (more precisely its moments) to be related to matrix
element of local operators as it happens for inclusive distributions. Indeed,
the operator definition of $f(\varepsilon)$ involves the ``maximally nonlocal
operator'' ${\cal E}(\vec n)$ that measures the density of the energy flow in
the direction of unit 3-vector $\vec n$. According to its definition ${\cal
E}(\vec n)$ acts on the final state of $N$ particles as ${\cal E}(\vec n)
\ket{N} = \sum_{i=1}^N k_i^0 \delta(\vec n-\vec k_i/k_i^0) \ket{N}$
and it can be expressed in terms of the tensor energy-momentum operator
\ci{ST,KOS}. Then, one uses \re{f} to get
\be
f(\varepsilon)=\vev{0|W^\dagger(0)\,\delta\lr{\varepsilon-\int d^3 n\,
(1-|\cos\theta_{\vec n}|)\,{\cal E}(\vec n)}\, W(0)|0}\,,
\lab{op}
\ee
where $\theta_{\vec n}$ is the angle between vector $\vec n$ and the thrust
axis, $\cos\theta_{\vec n}=n_3$. The detailed
properties of this function will be discussed elsewhere \ci{KS98}.

\section{Factorization of the thrust distribution}

The expression \re{nonPT} was found by neglecting perturbative soft gluons.
It is clear that they also contribute to the matrix element of Wilson lines
entering \re{fact} and modify the thrust distribution.  In order to combine
together perturbative and nonperturbative effects one has to introduce the
factorization scale $\mu$ and separate the contribution of soft gluons with the
momentum above and below this scale into perturbative Sudakov spectrum,
$\frac{d\sigma_{\PT}}{dt}$, and nonperturbative shape function,
$f(\varepsilon)$, respectively. Both quantities become functions of the IR
cut-off $\mu$ but the thrust distribution is $\mu-$independent.

For the inclusive distributions the above procedure can be performed using
the operator product expansion. For the thrust distribution the OPE is not
valid and we apply instead the infrared renormalon analysis. To this end
we perform perturbative calculation of \re{fact} by summing over the final
states $\ket{N}$ of multiple perturbative soft gluon radiation and identify
the ambiguities of resummed perturbative expressions that can be attributed
to nonperturbative contribution. One finds that thanks to the nonabelian 
exponentiation of the Wilson lines, the perturbative soft gluon 
contribution exponentiates in the Laplace transform of the distribution 
\ci{KS94}
\be
\vev{\e^{-\nu t}}\equiv
\int_0^{t_{\rm max}}dt \e^{-\nu t} \frac1{\sigma_{\rm tot}}
\frac{d\sigma}{dt}
=\exp\lr{-S(\nu)}
\lab{Lap}
\ee
with the leading term in the exponent of the following form
\be
S(\nu)=2\int_0^1\frac{du}{u}\lr{1-\e^{-u\nu}}
\int_{u^2Q^2}^{uQ^2}\frac{dk_\perp^2}{k_\perp^2}
\Gamma_{\rm cusp}(\as(k_\perp))
\lab{S}
\ee
and $\Gamma_{\rm cusp}(\as)=\frac{4\as}{3\pi}+\CO(\as^2)$ being a universal 
cusp anomalous dimension.

We observe that since the singularities of the coupling constant affect
the integration over transverse momenta of soft gluons in \re{S}
the Sudakov form factor $S(\nu)$ suffers from infrared renormalon ambiguities.
They originate from soft gluons with the energy of order $\Lambda_{\rm
QCD}$ whose contribution should be separated into nonperturbative function
$f(\varepsilon)$. Namely, introducing the cut-off $\mu$ on the value of
transverse gluon momenta in \re{S} we may split $S(\nu)$ into the
sum of two terms. The term with $k_\perp^2 > \mu^2$ defines the
perturbative contribution to the exponent, $S_\PT(\nu)$, which in turn
allows to find the perturbative spectrum $\frac{d\sigma_\PT(t;\mu)}{dt}$
through the inverse Laplace transformation \re{Lap}. The second term with
$k_\perp^2 < \mu^2$ should be absorbed into the definition of the
nonperturbative function \re{f}. In this case, changing the order of
integration in \re{S} one expands the integral in powers of $\nu$ and absorbs
the ambiguous integrals $\int_0^\mu dk_\perp k_\perp^{n-1}\Gamma_{\rm cusp}
(\as(k_\perp))$ into the definition of new nonperturbative scales
$\lambda_n(\mu)$. Substituting \re{nonPT} into \re{Lap}
one finds the following consistency condition
\be
\int_0^\infty d\varepsilon \e^{-\nu\varepsilon/Q} f(\varepsilon;\mu)
=\exp\lr{-\sum_{n=1}^\infty \frac1{n!}
(\nu/Q)^n \lambda_n(\mu)+ \CO(\nu/Q^2)}\,.
\lab{sum}
\ee
Although we can not draw any conclusions about the absolute value of the
scales $\lambda_n$, their $\mu-$dependence is of perturbative origin and
it can be determined as
\be
\mu\frac{d \lambda_n(\mu)}{d\mu}= {4 (-)^{n+1}}{n^{-1}}\mu^n \Gamma_{\rm
cusp}(\as(\mu))\,.
\lab{RG}
\ee
Since the parameter $\nu$ is conjugated to the
thrust $t$ we neglected in \re{sum} the corrections $\sim \nu/Q^2$ and replaced
the upper limit of $\varepsilon-$integration,
$\varepsilon_{\rm max}=t_{\rm max}Q$, by $\infty$.

The fact that the infrared renormalons contribute additively to the exponent
of \re{Lap} implies that the Laplace transform of the thrust distribution is
factorized into the product of perturbative and nonperturbative terms \ci{KS94}
\be
\vev{\e^{-\nu t}} = \vev{\e^{-\nu t}}_\PT \times
\int_0^\infty d\varepsilon \e^{-\nu\varepsilon/Q} f(\varepsilon;\mu)
\lab{exp}
\ee
where $\vev{\e^{-\nu t}}_\PT$ is calculated as a mean value with respect to the
perturbative distribution $\frac{d\sigma_\PT}{dt}$. Integrating the both
sides of this relation with respect to $\nu$ with an appropriate weight we 
obtain the factorized expression for the radiator function
$R(\tau)\equiv \frac1{\sigma_{\rm tot}} \int_0^\tau dt\,\frac{d\sigma}{dt}
=\vev{\theta(\tau-t)}$
\be
R(t)=\int_0^{tQ} d\varepsilon\, f(\varepsilon;\mu)
R_\PT\lr{t-\frac{\varepsilon}{Q};\mu}\,,
\lab{R}
\ee
where the upper limit of integration follows from the condition
$R_\PT(t)=0$ for $t < 0$.  
Thus, the net effect of incorporating nonperturbative corrections (in
the leading $1/(Q^2t)-$order) amounts to the $1/Q-$shift of perturbative
radiator function smeared with the shape function.

To see how \re{R} resums both perturbative and nonperturbate corrections,
one expands the radiator $R_\PT\lr{t-\frac{\varepsilon}{Q};\mu}$ in powers 
of $1/Q$
\be
R(t)=R_\PT(t) - \frac{\vev{\varepsilon}}{Qt} R_{\PT}'(t) +
\frac{\vev{\varepsilon^2}}{2(Qt)^2}[R_{\PT}''(t)-R_{\PT}'(t)] + \ldots \,,
\lab{R-exp}
\ee
where prime denotes the logarithmic derivative with respect to $t$.  Here,
the leading term $R_{\PT}(t)$ gives the resummed perturbative Sudakov
expression which is different from \re{PT} due to additional dependence on 
the IR cut-off $\mu$. The terms with the derivatives of $R_{\PT}$ 
generate the series of power corrections accompanied by the set of 
nonperturbative $\mu-$dependent dimensionful parameters $\vev{\varepsilon^n}$ 
that can be expressed in terms of the scales $\lambda_k$ introduced in 
\re{sum} as
\be
\vev{\varepsilon^n} =
\int_0^\infty d\varepsilon\,\varepsilon^n f(\varepsilon;\mu) \,,\qquad
\vev{\varepsilon}=\lambda_1
\,,\quad
\vev{\varepsilon}^2-\vev{\varepsilon^2}=\lambda_2
\,,\quad ...
\lab{mom}
\ee
Finally, differentiating the both sides of \re{R} with respect to $t$ and
taking into account that $R_{\PT}(t;\mu)=\theta(t)\int_0^t dt
\frac{d\sigma_{\PT}(t;\mu)}{dt}$ we find the thrust distribution
\be
\frac1{\sigma_{\rm tot}}\frac{d\sigma(t)}{dt}
=Q f(Qt;\mu) R_{\PT}(0;\mu)
+\int_0^{Qt} d\varepsilon\,f(\varepsilon;\mu)
\frac{d\sigma_{\PT}(t-\frac{\varepsilon}{Q};\mu)}{dt}\,.
\lab{thrust}
\ee
Here, the two terms in the r.h.s.\ have the following interpretation.
The first term dominates at small $t$ and corresponds to the
situation when the perturbative real soft gluon radiation is washed out
due to the IR cut-off and the final state consists entirely of
nonperturbative radiation described by the shape function. In
comparison with \re{nonPT} one gets the additional Sudakov factor
$R_{\PT}(0;\mu)$ that takes into account the perturbative contribution
of virtual soft gluons with the energy above the cut-off $\mu$. This
factor rapidly vanishes as one decreases the value of $\mu$ allowing
more perturbative virtual gluons to be emitted. The second term in
\re{thrust} describes the smearing of the perturbative Sudakov spectrum
by nonperturbative corrections. As an illustration of \re{thrust}, we 
depicted on Fig.~1 the contribution of both terms to the thrust 
distribution at the center-of-mass energy $Q=35\,{\rm GeV}$. 
\begin{figure}[!ht]
\centerline{\epsfysize=10cm\epsfxsize=12cm
\epsffile{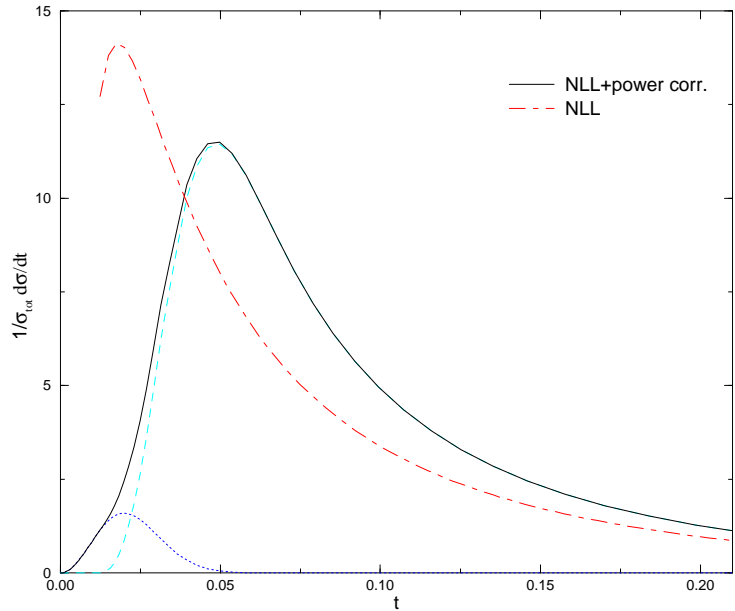}}
\vspace*{-7mm}
\caption{The prediction \re{thrust} for the thrust distribution
at $Q=35\,{\rm GeV}$. The dotted and dashed lines describe the
first and the second term in the r.h.s.\ of \re{thrust}, respectively,
and the solid line is the sum of both. The dot-dashed line denotes
the perturbative Sudakov spectrum 
$\frac{d\sigma_{\PT}(t;\mu=0)}{dt}$ in the NLL
approximation.}
\end{figure}

In contrast with the heavy meson decay
where nonperturbative corrections extend the perturbative spectrum
beyond the perturbative end-point due to interaction of the heavy
quark in the initial state with the light component of the meson 
\ci{BSUV,N}, the nonperturbative corrections to the thrust distribution 
have just an opposite effect. They shift the perturbative spectrum inside the
perturbative window $0 < t < t_{\rm max}$ and describe an ``evaporation''
of the energetic jets in the final state due to emission of soft 
gluons.

\section{Nonperturbative ansatz for the shape function}

The shape function is a new distribution that resums nonperturbative
corrections to the thrust distribution. Although its explicit form can
not be extracted from our analysis we could use the renormalon inspired
sum rules \re{sum} to study its general properties.

Expanding the both sides of \re{sum} in powers of $1/Q$ we verify that
in accordance with its operator definition \re{op} the function
$f(\varepsilon)$ does not depend on the large scale $Q$ and its first few
moments are given by \re{mom}. The shape function depends on the factorization
scale $\mu$ and one may apply the renormalization group equations \re{RG} to 
find its evolution with $\mu$.

Using \re{mom} one may formally write the shape function in the form
of the distribution as a series in $\delta-$function and its
derivatives
\be
f(\varepsilon;\mu)=\delta(\varepsilon-\lambda_1)
+ 0\cdot \delta'(\varepsilon-\lambda_1)
-\frac12\lambda_2 \delta''(\varepsilon-\lambda_1)+ ...
\lab{f-exp}
\ee
Its substitution into \re{thrust} yields a series that is equivalent to
the expansion of the radiator \re{R-exp}. It is convergent however only for
the values of the thrust $\Lambda_\QCD/Q \ll t < t_{\rm max}$ on the tail
of the Sudakov spectrum $\frac{d\sigma_\PT}{dt}$ where the perturbative
distribution is a slowly varying function of $t$. In this range of $t$,
keeping only the first term in the r.h.s.\ of \re{f-exp}, one finds that the
leading power corrections simply renormalize the thrust variable by
generating the shift of the perturbative
spectrum \ci{KS94}
\be
\frac1{\sigma_{\rm tot}}\frac{d\sigma(t)}{dt}
=\frac{d\sigma_{\PT}(t-\frac{\vev{\varepsilon}}{Q})}{dt}\,.
\lab{t-large}
\ee
This result is a general property of the leading $1/Q-$power corrections
to different event shapes and it is an immediate consequence of the
exponentiation of soft gluon contribution. The prediction
\re{t-large} has been found to be in a good agreement with the data 
\ci{DW}.

For the values of the thrust $t =\CO(\Lambda_\QCD/Q)$ one needs to know
the explicit form of the shape function. In this case, one relies on the
particular ansatz for $f(\varepsilon)$ that can be inspired by  
different model considerations. As the simplest form of $f(\varepsilon)$ we
choose the following one
\be
f(\varepsilon)=\lr{\frac{\varepsilon}{\Lambda}}^{a-1}
\exp\lr{-\frac{\varepsilon^2}{\Lambda^2}} \frac2{\Lambda\Gamma(\frac{a}2)}\,.
\lab{ans}
\ee
This function depends on two parameters: dimensionless exponent $a$
controlling how fast the function vanishes at the origin and dimensionfull
scale $\Lambda$ defining the interval of energies on which the function
is localized. The shape function is peaked around $\varepsilon_{\rm max}=
\Lambda\sqrt{\frac{a-1}2}$ and it rapidly vanishes for
$\varepsilon > \varepsilon_{\rm max}$.

There are additional constraints that we may impose on the shape function.
They follow from the analysis of nonperturbative corrections to the mean
value of the thrust and its higher moments. Indeed, expanding the both
sides of \re{exp} in powers of $\nu$ one gets
\be
\vev{t}=\vev{t}_{_{\rm PT}} + \frac{\vev{\varepsilon}}{Q}\,,\qquad
\vev{t^2}-\vev{t}^2 = \vev{t^2}_{\PT}-\vev{t}^2_{\PT} +
\frac{\vev{\varepsilon^2}-\vev{\varepsilon}^2}{Q^2}\,,\quad ...
\lab{vev}
\ee
These relations are valid up to corrections due to contribution to
the thrust of the final states with $n\ge 3$ jets. Using \re{vev} and
\re{mom} we may relate the parameters of the shape function to the mean
value of the first few moments of the thrust. In particular, applying
the results of \ci{evi} on parameterization of $1/Q-$corrections
to the mean value of the thrust we get 
$$
\vev{t}-\vev{t}_{\PT}=
\frac1{Q} \int_0^\infty d\varepsilon \,
\varepsilon f(\varepsilon) = \frac{0.8\ \rm{GeV}}{Q}
$$
Substituting \re{ans} into this relation one still has a freedom
in choosing the single parameter $a$ that we define as
\be
a=3\,,\qquad \Lambda=0.7 \ {\rm GeV}\,.
\lab{para}
\ee
The shape function corresponding to these values of parameters
is shown in Fig.~2. One should stress that the explicit form the shape
function \re{ans} as well as the values of the parameters $a$ and $\Lambda$
are related to the particular choice of the factorization scale $\mu$ to be
specified later on and they do not depend on the center-of-mass energy $Q$.

\begin{figure}[ht]
\centerline{\epsfysize=8cm\epsfxsize=10cm
\epsffile{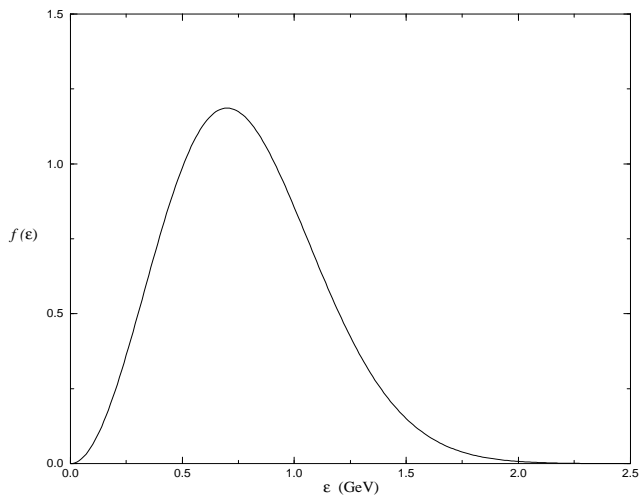}}
\vspace*{-7mm}
\caption{The nonperturbative ansatz for shape function.}
\end{figure}

Let us now consider the perturbative spectrum $\frac{d\sigma_{\PT}(t;\mu)}
{dt}$. The contribution of perturbative soft gluons with the energy above
the IR cut-off $\mu$ exponentiates in the Laplace transform
\re{Lap} and
$S_\PT(\nu)$ is given by \re{S} with the additional condition $k_\perp^2 >
\mu^2$ imposed on the $k_\perp-$integration. The important difference with 
the known results for resummed Sudakov spectrum \ci{CTTW} is in the 
additional $\mu-$dependence of $\frac{d\sigma_{\PT}(t;\mu)}{dt}$.
Nevertheless, one may expand following \ci{CTTW} the exponent $S_{\PT}(\nu)$ 
in powers of $\as(Q)$ and separate large
logarithmic terms $\as^n L^{m}$ $(m \le n+1)$ which due the presence
of additional scale appear of two kinds, $L=\ln(Q/\mu)$ and $L=\ln(1/\nu)$.
Then, in the NLL approximation that takes into account all terms
$\as^n L^{n+1}$ and $\as^n L^n$ in $S_{\PT}(\nu)$ one may replace
$1-\e^{-u\nu}\stackrel{_{\rm NLL}}{=}\theta(u-\e^{-\gamma_{_{\rm E}}}
/\nu)$ in \re{S} and find that due to the condition $k_\perp > \mu$
the function $S_{\PT}(\nu)$ has different behaviour depending on
the value of $\nu$, or equivalently the thrust $t$. Finally, one
performs the inverse Laplace transformation and obtains
the radiator function in the NLL approximation as
\be
R_{\PT}(t;\mu)\stackrel{_{\rm NLL}}{=}\frac{\exp\lr{-S_{\PT}(t_0/t)}}
{\Gamma(1-S_{\PT}'(t_0/t))}\,,
\qquad
S_{\PT}'(t_0/t)= \frac{\partial S_{\PT}(t_0/t)}{\partial \ln t}\,,
\lab{R-NLO}
\ee
with $t_0=\e^{\gamma_{_{\rm E}}}=1.780...$. Examining this expression one
finds that for $t/t_0>\mu/Q$ the radiator $R_{\PT}(t;\mu)$ rapidly vanishes
at small $t$ and it does not depend on the IR cut-off $\mu$. Thefefore
for these values of $t$ it coincides with the known expression from \ci{CTTW}.
For $\mu^2/Q^2 < t/t_0 < \mu/Q$ the radiator $R_{\PT}(t)$ starts to depend on
the IR cut-off $\mu$ and, as a consequence, its decrease at small $t$ is
slowed down. For $0< t/t_0 < \mu^2/Q^2$ the radiator is $t-$independent.

To simplify numerical calculations of the spectrum \re{thrust} we ignore the
difference between $R_{\PT}(t=t_0\mu/Q)$ and $R_{\PT}(t=t_0\mu^2/Q^2)$ by 
choosing the factorization scale $\mu$ to be small but within the
applicability range of the NLL approximation \ci{CTTW}, 
$2\beta_0\as(Q) \ln(Q/\mu t_0) < 1$. In this case, one may
approximate the radiator \re{R-NLO} as
\be
R_{\PT}^{\rm app}(t;\mu)=\theta(t/t_0-\mu/Q) R_{\PT}(t)
+ \theta(t)\theta(\mu/Q-t/t_0) R_{\PT}(\mu/Q)\,,
\lab{app}
\ee
where $R_{\PT}(t)$ is the known expression for the radiator in the
NLL approximation \ci{CTTW} improved by higher order corrections in the
$\ln R-$matching scheme. We choose the two free parameters,
the factorization scale and the fundamental QCD scale as
$$
\mu=0.750 \ {\rm GeV}\,,\qquad
\Lambda_{\QCD{}_{;}\MS}^{(5)}=0.250\,{\rm GeV}\,.
$$
Differentiating the expression \re{app} with respect to $t$ we
obtain the perturbative spectrum $\frac{d\sigma_{\PT}(t;\mu)}{dt}$
that starts at $t=t_0\mu/Q$
and extends to $t=t_{\rm max}$.

One should notice that if instead of \re{app} we would have used for
$t/t_0\le \mu/Q$ the expression for the radiator $R_{\PT}(t;\mu)$
in the NLL order, \re{R-NLO}, then the perturbative spectrum will have
discontinuities at $t/t_0=\mu/Q$ and $t/t_0=\mu^2/Q^2$. The reason for this 
is that
$\frac{d\sigma_{\PT}(t;\mu)}{dt}=\frac{R_\PT(t;\mu)}{dt}$ involves the second
order derivative of the Sudakov form factor, $S_{\PT}''$, that, as it can be
seen from \re{S}, is not enhanced by large logarithmic corrections and
therefore can not be approximated by the NLL result in the transition
regions around these two points.

Finally, let us compare the QCD prediction \re{thrust} with available data
on the thrust distribution in the interval $0<t<0.33$ at different energies
$14 < Q/{\rm GeV} < 161$. Fig.~3 shows the comparison with the data at $Q=91.2\,{\rm GeV}$,
where the most accurate experimental data are available.
The combined fit for various center-of-mass energies \ci{data} is
shown in Fig.~4. We would like to stress that for different values of $Q$
we use the {\it same\/} ansatz for the shape function, \re{ans}, with the
parameters defined in \re{para}. As a nontrivial test of \re{thrust} we
observe that the theoretical curves reproduce well the $Q-$dependence of
the data in the end-point region.

\begin{figure}[ht]
\centerline{\epsfysize=10cm\epsfxsize=12cm
\epsffile{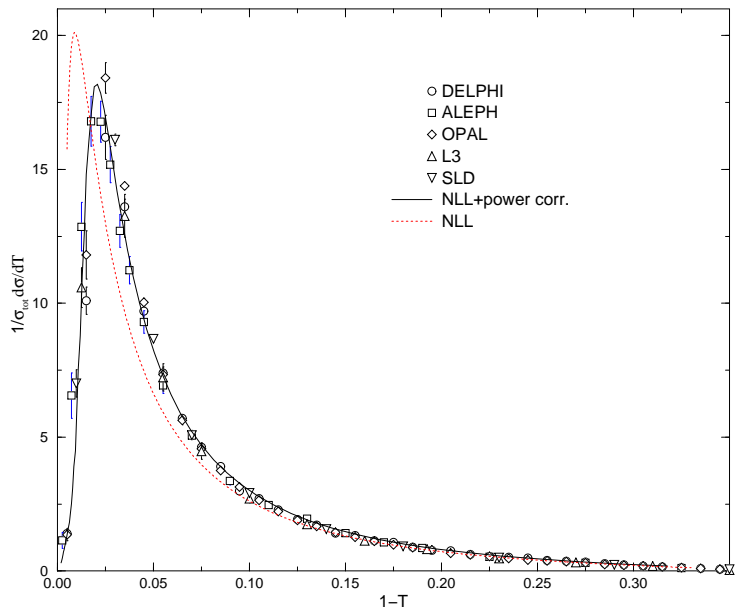}}
\vspace*{-7mm}
\caption{The comparison of the data with the QCD prediction
for the thrust distribution \re{thrust} at $Q=91.2\,{\rm GeV}$}
\end{figure}

\begin{figure}[ht]
\centerline{\epsfysize=10cm\epsfxsize=12cm
\epsffile{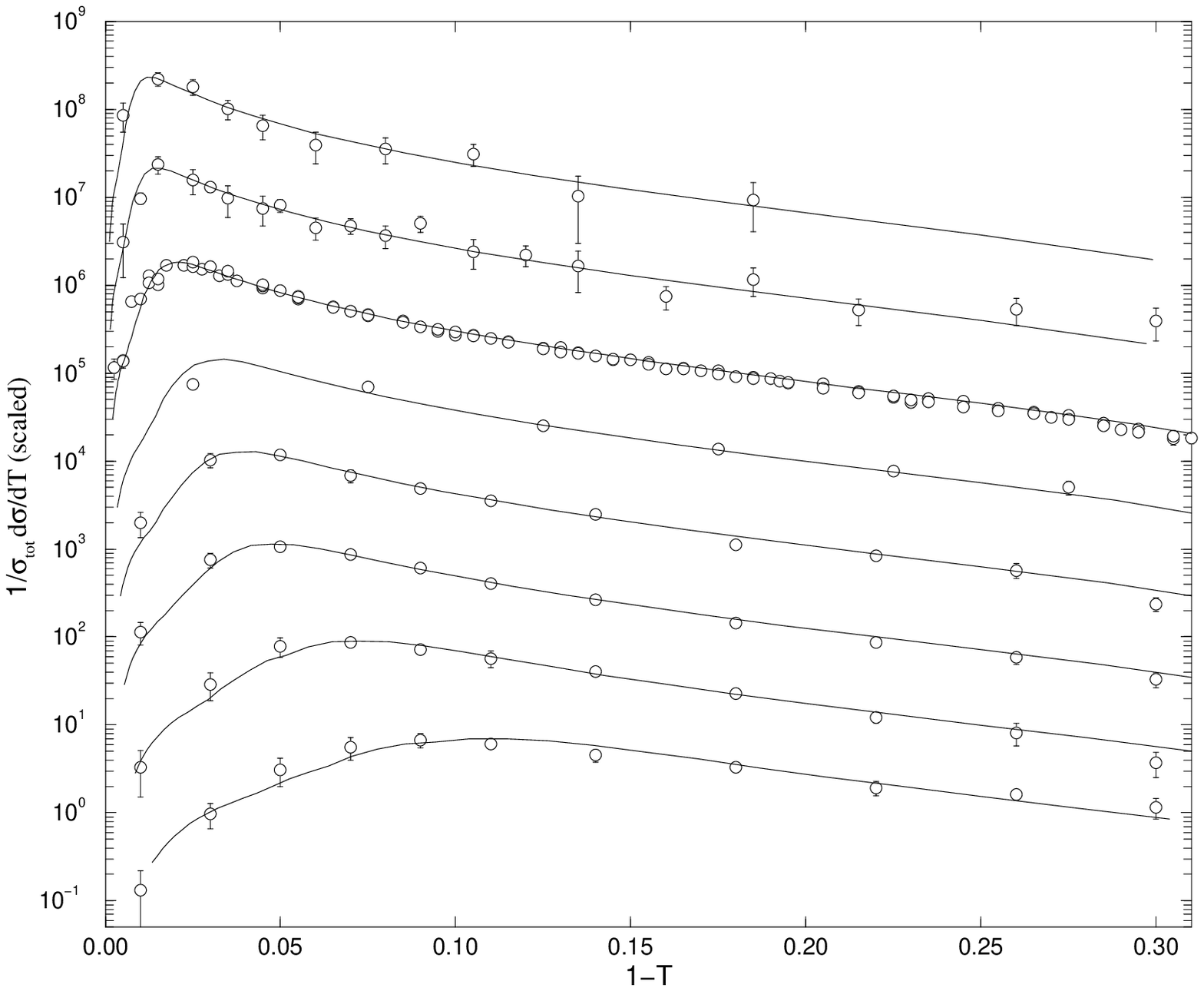}}
\vspace*{-7mm}
\caption{The comparison of the data with the QCD prediction for the 
thrust distribution at different energies (from bottom to top):
$Q/{\rm GeV}=14\,,22\,,35\,,44\,,55\,,91\,,133\,,161.$ \ci{data}}
\end{figure}

\section{Conclusions}

We have study the power corrections to the thrust 
differential distribution in the end-point region of $t$. Due to enhancement 
of soft gluon contribution, the spectrum is affected by large perturbative 
Sudakov and 
nonperturbative $1/(Qt)-$corrections that need to be resummed. Using universal 
properties of soft gluon radiation we have argued that the resummed leading 
power corrections are described by the shape function which is a new 
nonperturbative distribution independent on the center-of-mass energy $Q$
measuring the energy flow in the final state. The 
thrust distribution is given by the convolution of the shape function and the 
perturbative Sudakov spectrum each depending on the factorization scale $\mu$. 
For the values of the thrust $t > \mu/Q$ the power corrections generate the shift 
of the perturbative spectrum, \re{t-large}. In this case, the thrust 
distribution is not sensitive to the particular form of the shape function
but only to its first moment $\vev{\varepsilon}$. In contrast, for
${\Lambda_\QCD^2}/{Q^2} \ll t < {\mu}/{Q}$ it is the shape function that
governs the end-point behaviour of the spectrum. Choosing the simplest
ansatz \re{ans} for this function and using the $\ln R-$matched expression
for the perturbative spectrum \re{app} we have found that our prediction 
\re{thrust} provides a good description of the data in a wide range of
energies.

Analysing the power corrections to the event shapes one should
identify universal nonperturbative quantities that describe
the hadronization effects in $\e^+\e^-$ annihilation. According to
\re{op} the shape function depends on the definition of the thrust variable
and considering the power corrections to other event shapes like heavy mass
jet or energy-energy correlations one gets  the expressions for
the shape functions as well as the factorized expressions for the
distributions different from \re{op} and \re{thrust}. However, taking the
moments of the shape function, $\int d\varepsilon \,\varepsilon^N
f(\varepsilon)$, one finds that for various event shapes they are expressed
in terms of the same {\it universal\/} distribution
$\vev{0| W^\dagger(0){\cal E}(\vec n_1) ... {\cal E}(\vec n_N) W(0)|0}$
that measures the energy flow in the final state in the directions
specified by unit 3-vectors $\vec n_1$, $...$, $\vec n_N$. This object
deserves additional studies \ci{KS98}.

\section*{Acknowledgments}

I am grateful to A.B.~Kaidalov and G.~Sterman for illuminating discussions.
This work was supported in part by the NSF/CNRS grant.

\bb{99}
\bi{evi}  DELPHI Coll., Z. Phys. C73 (1997) 22;
\\        ALEPH Coll., Contribution to EPS-HEP97, 
          Jerusalem 19-26 Aug. 1997, abstract 610. 
\\        D. Wicke, Nucl. Phys. Proc. Suppl. 64 (1998) 27. 
\\        P.A. Movilla Fernandez, et. al. and the JADE Coll., 
          Eur. Phys. J. C1 (1998) 461.
\\        O. Biebel, Nucl. Phys. B, Proc. Suppl. 64 (1998) 22.
\\        J.M. Campbell, E.W.N. Glover and C.J. Maxwell, Durham preprint 
          DTP-98-8, hep-ph/9803254.
\bi{W}    B.R. Webber, Phys. Lett. B339 (1994) 148;
\\        Yu.L. Dokshitzer and B.R. Webber, Phys. Lett. B352 (1995) 451.	  
\bi{KS94} G.P. Korchemsky and G.Sterman, Nucl. Phys. B437 (1995) 415; in
          {\it QCD and High Energy Hadronic Interactions}, proceedings of the 
	  30th Rencontres de Moriond, Les Arcs, Savoie, France, 18-25 March,
          ed. J. Tran Thanh Van (Editions Frontieres, Gif-sur-Yvette, 1995), 
	  p.383; hep-ph/9505391. 
\bi{AZ}   R. Akhoury and V. Zakharov, Phys. Lett. B357 (1995) 646;
          Nucl. Phys. B465 (1996) 295.
\bi{BB}   M. Beneke and V.M. Braun, Nucl. Phys. B454 (1995) 253.
\bi{NS}   P. Nason and M.H. Seymour, Nucl. Phys. B454 (1995) 291.	
\bi{DMW}  Yu.L. Dokshitzer, G. Marchesini and B.R. Webber, Nucl. Phys. B469
          (1996) 93;
\\        Yu.L. Dokshitser, A. Lucenti, G. Marchesini and G.P. Salam, 
          Nucl. Phys. B511 (1998) 396; J. High Energy Phys. 5 (1998) 3.  
\bi{DIS}  G. Sterman, Nucl. Phys. B281 (1987) 310;
\\        S. Catani and L. Trentadue, Nucl. Phys. B327 (1989) 323; 
          B353 (1991) 183;
\bi{WL}   G.P. Korchemsky, Mod. Phys. Lett. A4 (1989) 1257; 
\\        G.P. Korchemsky and G. Marchesini, Phys. Lett. B313 (1993) 433.
\bi{BSUV} I.I. Bigi, M.A. Shifman, N.G. Uraltsev and A.I. Vainshtein,  
          Int. J. Mod. Phys. A9 (1994) 2467;
\\        R.D. Dikeman, M. Shifman and N.G. Uraltsev, Int. J. Mod. Phys. 
          A11 (1996) 571. 	  
\bi{N}    M. Neubert, Phys. Rev. D49 (1994) 4623, 3392.
\bi{GK}   G.P. Korchemsky and G. Sterman, Phys. Lett. B340 (1994) 96;
\\        A.G. Grozin and G.P. Korchemsky, Phys. Rev. D53 (1996) 1378.
\bi{CTTW} S. Catani, L. Trentadue, G. Turnock and B.R. Webber,
          Nucl. Phys. B407 (1993) 3.
\bi{ST}   N.A. Sveshnikov and F.V. Tkachev, Phys. Lett. B382 (1996) 403; 
\\        P.S. Cherzor and N.A. Sveshnikov, hep-ph/9710349.	  
\bi{KOS}  G.P. Korchemsky, G. Oderda and G. Sterman, in proceedings of the
          5th International Workshop ``Deep Inelastic Scattering and QCD'', 
	  ed.\ J. Repond and D. Krakauer, AIP Conf. Proc. No.407, 
	  Woodbury, NY, 1997, p.988; hep-ph/9708346.
\bi{KS98} G.P. Korchemsky and G. Sterman, in preparation.
\bi{DW}   Yu.L. Dokshitzer and B.R. Webber, Phys. Lett. B 404 (1997) 321.
\bi{data} ALEPH Coll., Phys. Lett.  B284 (1992) 163; Z. Phys.  C55 (1992) 209; 
          CERN-PPE-96-186;
          AMY Coll., Phys. Rev. Lett.  62 (1989) 1713; Phys. Rev. D41 (1990) 2675;
  	  CELLO Coll., Z. Phys.  C44 (1989) 63;
 	  DELPHI Coll., Z. Phys. C59 (1993) 21; Z. Phys. C73 (1996) 11, 229;
 	  HRS Coll.,   Phys. Rev.  D31 (1985) 1;
 	  JADE Coll.,   Z. Phys.  C25 (1984) 231; Z. Phys.  C33 (1986) 23;
          L3 Coll.,   Z. Phys.  C55 (1992) 39;
 	  Mark II Coll.,   Phys. Rev.  D37 (1988) 1; Z. Phys.  C43 (1989) 325;
 	  MARK J Coll.,   Phys. Rev. Lett.  43 (1979) 831;
 	  OPAL Coll., Z. Phys. C55 (1992) 1; Z. Phys. C59 (1993) 1; 
	  Z. Phys. C72 (1996) 191; Z. Phys. C75 (1997) 193; 
 	  PLUTO Coll.,   Z. Phys.  C12 (1982) 297; 
 	  SLD Coll.,   Phys. Rev.  D51 (1995) 962;
 	  TASSO Coll.,   Phys. Lett.  B214 (1988) 293;
   	  Z. Phys.  C45 (1989) 11;
   	  Z. Phys.  C47 (1990) 187; 
 	  TOPAZ Coll.,  Phys. Lett.  B227 (1989) 495;
   	  Phys. Lett.  B278 (1992) 506; 
   	  Phys. Lett.  B313 (1993) 475
\eb

\end{document}